\documentclass[aps,english,prb,twocolumn,superscriptaddress]{revtex4}
\usepackage{graphicx}
\usepackage{amssymb}
\usepackage{babel}

\begin{document}

\title{The response of mechanical and electronic properties of graphane to the elastic strain}

\author{M. Topsakal}
\affiliation{UNAM-Institute of Materials Science and Nanotechnology, Bilkent University, Ankara 06800, Turkey}
\author{S. Cahangirov}
\affiliation{UNAM-Institute of Materials Science and Nanotechnology, Bilkent University, Ankara 06800, Turkey}
\author{S. Ciraci}
\affiliation{UNAM-Institute of Materials Science and Nanotechnology, Bilkent University, Ankara 06800, Turkey}
\affiliation{Department of Physics, Bilkent University Ankara 06800, Turkey}

\date{\today}

\begin{abstract}
Based on first-principles calculations, we resent a method to
reveal the elastic properties of recently synthesized monolayer
hydrocarbon, graphane. The in-plane stiffness and Poisson's ratio
values are found to be smaller than those of graphene, and its
yielding strain decreases in the presence of various vacancy
defects and also at high ambient temperature. We also found that
the band gap can be strongly modified by applied strain in the
elastic range.

\end{abstract}

\maketitle

Two dimensional (2D) monolayer honeycomb structures of
graphene,\cite{gra1,gra2} BN,\cite{bn1} and
silicon\cite{silicene} offer remarkable properties and are
promising materials for future applications. Honeycomb structure
of graphene with $sp^{2}$ bonding underlies the unusual mechanical
properties providing very high in-plane strength. Graphene and its
rolled up forms, carbon nanotubes are among the strongest and
stiffest materials yet discovered in terms of tensile strength and
elastic modulus.\cite{strongest1,strongest2} Graphane,  another
member of honeycomb structures was theoretically
predicted\cite{sofo} and recently  synthesized by exposing
graphene to hydrogen plasma discharge.\cite{graphane_science} Here
each carbon atom being bonded to one hydrogen atom is pulled out
from the graphene plane and hence whole structure is buckled.
Instead of being a semimetal like graphene, graphane is a wide
band gap semiconductor and can attain permanent magnetic moment
through hydrogen vacancies.\cite{hasan}

In this work, we revealed the relevant elastic constants of
graphane using strain energy calculations in the harmonic elastic
deformation range and compared them with those calculated for
other honeycomb structures. We also found that in the presence of
hydrogen vacancy and carbon+hydrogen divacancy, its yielding
occurs at smaller strains. Furthermore, its band gap first
increases then decreases steadily with the increasing applied
strain. We believe that our predictions are relevant for the
current research focused on the electronic properties of honeycomb
structures under strain.\cite{neto,topsakal_kopma}

First-principles plane wave calculations are carried out within
density functional theory using PAW potentials.\cite{paw}
The exchange correlation potential is approximated by generalized
gradient approximation (GGA) using PW91 functional. A plane-wave
basis set with kinetic energy cutoff of 450 eV is used. All atomic
positions and lattice constants are optimized by using the
conjugate gradient method, where the total energy and atomic
forces are minimized. Interactions  between adjacent graphane
layers in supercell geometry is hindered by a large spacing of
$\sim$10 \AA. To correct the energy bands and band gap values
obtained by GGA, frequency-dependent $G_{0}W_{0}$ calculations are
carried out. $G_{0}W_{0}$ corrections are obtained by using
(12x12x1) \textbf{k}-points in the Brillouin zone (BZ), 400 eV cut-off potential for
$G_{0}W_{0}$, 160 bands and 64 frequency grid points. All
numerical calculations are performed by using VASP
package.\cite{vasp,gw}

\begin{figure}
\includegraphics[width=7.9cm]{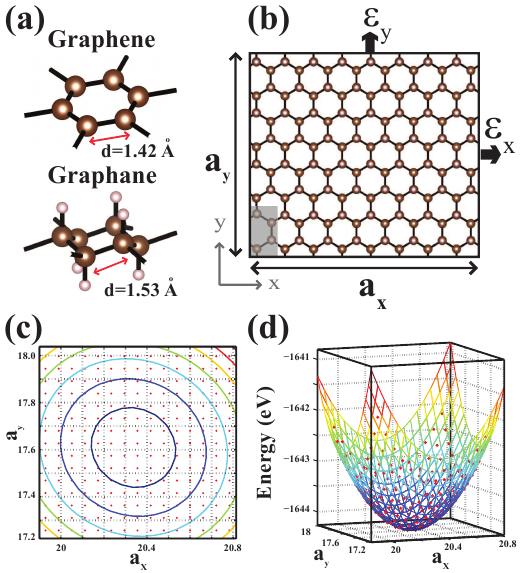}
\caption{(Color Online) (a) Schematic representation of the atomic
structure of graphene and graphane. (b) (8$\times$4) rectangular
supercell configuration of the system containing 128 C-H pairs
used for the calculation of the elastic constants. $a_{x}$ and
$a_{y}$ are the lattice constants of the supercell in $x$- and
$y$-directions. Shaded region is the smallest unit cell. (c) The
mesh of data points $(a_{x}, a_{y})$ used for the total energy
calculations. The units are given in Angstroms. (d) The 3D plot of
$a_{x},a_{y}$ and corresponding total energy values. The red balls
are actual points and the lines are the fitted formula.}
\label{fig:1}
\end{figure}

The graphene has a 2D hexagonal unit cell with a lattice constant
of $\textit{a}=2.47 $ \AA. The C-C bond length is \textit{d}=1.42
\AA{} and all atoms lie in the same plane. Upon hydrogenation, the
lattice constant increases to 2.54 \AA~ and \textit{d} increases
to 1.53 \AA. Moreover, C-H bonds are 1.11 \AA{} and the amount of
buckling between the alternating carbon atoms in a hexagon is 0.46
\AA. Atomic configuration of graphene and graphane structures are
shown in Fig.~\ref{fig:1} (a).

The elastic properties of homogeneous and isotropic materials can
be represented by two independent constants, Young's modulus
\textit{Y} and Poisson's ratio $\nu$. Since the thickness of a
monolayer structure $h$ is ambiguous, the in-plane stiffness
\textit{C} is a better measure of the strength rather than Young's
modulus. Defining $A_{0}$ as the equilibrium area of the system,
the in-plane stiffness can be given as,
$\textit{C}=\frac{1}{A_{0}}(\frac{\partial^{2}E_{S}}{\partial\epsilon^{2}})$,
where $E_{S}$ is the strain energy calculated by subtracting the
total energy of the strained system from the equilibrium total
energy and $\epsilon$ is the uniaxial strain ($\epsilon = \Delta
a/ a$, $a$ being the lattice constant). The Poisson's ratio which
is the ratio of the transverse strain to the axial strain can be
defined straightforwardly as
$\nu$=-$\epsilon_{trans}$/$\epsilon_{axial}$.

\begin{figure}
\includegraphics[width=7.9cm]{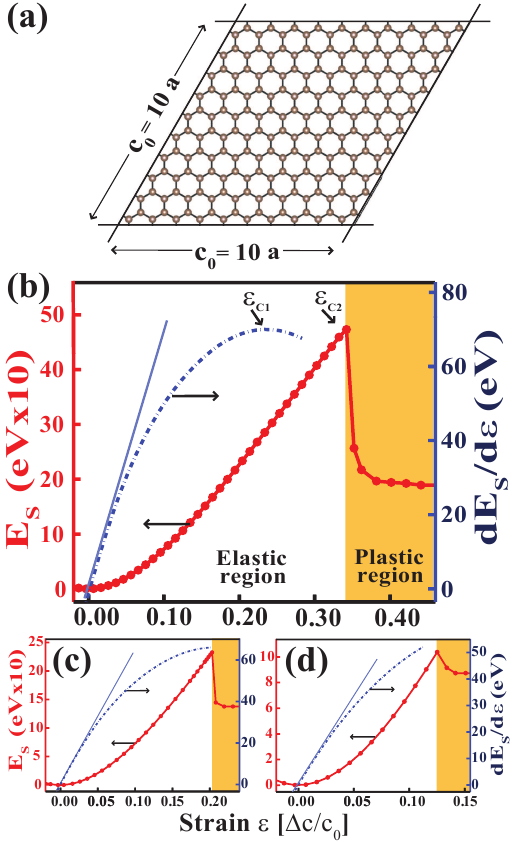}
\caption{(Color Online) Two dimensional graphane under uniform
expansion. (a) Initial atomic configuration in a (10$\times$10)
supercell treated with periodic boundary condition. (b) The
variation of strain energy $E_S$ and its derivative. The
orange/shaded region indicating the plastic range. Strains
corresponding to two critical points in the elastic range are
labeled as $\epsilon_{c_{1}}$ and $\epsilon_{c_{2}}$. (c) Similar
to (b) for a single H-vacancy in a (10$\times$10) supercell. (d)
For C+H-divacancy in a (10$\times$10) supercell. } \label{fig:2}
\end{figure}

For calculation of elastic constants of graphane, we consider
large supercell comprising 32 rectangular unit cells (8$\times$4).
The calculations are also repeated in (2$\times$1), (4$\times$2)
and (6$\times$3) supercells and the obtained results are almost
identical, since no reconstructions are observed in the system.
Fig.~\ref{fig:1} (b) shows the supercell used in the calculations.
$a_{x}$ and $a_{y}$ are the lattice constants of the supercell in
$x$- and $y$-directions in any strain condition. In the harmonic
region, $\textit{a}'s$ are varied with the strain values between
+/- 0.02. A grid data $(a_{x};a_{y})$ containing 225 points is
obtained as shown in Fig.~\ref{fig:1} (c). For each grid point,
the corresponding supercell is fully  optimized and its total
energy is calculated as shown in Fig.~\ref{fig:1} (d). By using
the least squares method, the data is fitted to the formula;
$E_{S}=a_{1}\epsilon_{x}^{2}+a_{2}\epsilon_{y}^{2}+a_{3}\epsilon_{x}\epsilon_{y}$
; where $\epsilon_{x}$ and $\epsilon_{y}$ are the small strains
along $x$- and $y$-directions in the harmonic region. As a result
of isotropy in the honeycomb symmetry, $a_{1}$ is equal to
$a_{2}$. The same equation can be obtained from elasticity
matrix\cite{nye} in terms of elastic stiffness constants, namely
$a_{1}=a_{2}= (h\cdot A_{0}/2) \cdot C_{11}$ ; $a_{3}=(h \cdot
A_{0}) \cdot C_{12}$. Hence one obtains Poisson's ratio $\nu$
which is equal to $C_{12}/C_{11}=a_{3}/2a_{1}$. Similarly, the
in-plane stiffness,  $C=h\cdot C_{11}\cdot(1-(C_{11}/C_{12})^{2}$)
= (2a$_{1}$-(a$_{3}$)$^{2}$/2a$_{1}$)/(A$_{0}$). The calculated
values of $C$ by using the present method for graphane, graphene,
BN, Si and SiC 2D honeycomb structures are, respectively, 243,
335, 267, 62 and 166 $(J/m^{2})$. Also the calculated Poisson's
ratios are 0.07, 0.16, 0.21, 0.30 and 0.29. Our calculated value
of the in-plane stiffness of graphene is in good agreement with
the experimental value\cite{strongest1} of 340$\pm$50 (N/m) and
justifies the reliability of our method. As seen from the
calculated values, the change of the bonding type from $sp^{2}$ to
$sp^{3}$ and buckling of the atoms in graphane structure makes it
27\% less stiffer than graphene. This difference can be used to
distinguish graphene and graphane materials. Also the Poisson's
ratio of graphane is almost half of the Poisson's ratio of
graphene, since the buckled structure of graphane reduces the
transverse contraction. Note that depending on their types and
concentrations the defects can alter the above elastic constants.
For example, a C$_2$H$_2$-vacancy for the structure in Fig.~\ref{fig:1} (b)
breaks the isotropy and can reduce $C$ by $\sim$12\% in a specific
direction. Hydrogen frustration\cite{graphane_science, flores} can
also be a crucial type of defect, which would affect $C$, since
the structure is locally compressed and A$_{0}$ is influenced.

We next consider the behavior of the system for higher values of
the strain ranging from -0.02 to 0.45 in uniform expansion. For
this purpose, we preferred a fully symmetric hexagonal lattice
with well defined high symmetry points in the BZ. Again the
calculations are performed in a large (10$\times$10) supercell as
shown in Fig.~\ref{fig:2} (a). The harmonic region can be taken
between $-0.02<\epsilon < 0.02$ and it is followed by an
anharmonic region where higher order terms are not negligible in
the strain energy equation. The anharmonic region is followed by a
plastic region where irreversible structural changes occur in the
system and it transforms into a different structure after the
yielding point. Fig.~\ref{fig:2} (b) is the plot of strain energy
$E_{S}$ and its derivative ($dE_{S}(\epsilon)/d\epsilon$) with
respect to the applied strain. Two critical strain values can be
deduced from the plots. The first one, $\epsilon_{c_{1}}$, is the
point where the derivative curve attains its maximum value and
then starts to decrease. It occurs nearly at $\epsilon=0.23$,
where the C-C bond length is around 1.87 \AA. This means that for
$\epsilon>\epsilon_{c_{1}}$, the  structure can be expanded under
smaller tensions. The phonon frequencies, we calculated by using
the force constant method\cite{alfe} are all positive throughout
the BZ for $\epsilon<\epsilon_{c_{1}}$, but the frequencies of
longitudinal acoustic modes start to become imaginary for
$\epsilon>\epsilon_{c_{1}}$, indicating an instability of 2D
graphane under uniform expansion beyond $\epsilon_{c_{1}}$. Such
phenomena is known as ``phonon
instability''\cite{instability1,instability2}, where  phonon
frequencies $\Omega_{n}$(\textbf{k}), get imaginary  for specific
wave vector \textbf{k} and branch index \textit{n}. A detailed
discussion can be found in Ref. \onlinecite{instability1} and the
references therein. Liu \textit{et al.}\cite{instability1}
calculated the critical strain values for graphene as 0.194 and
0.266 for uniaxial tension in zigzag (x-) and armchair (y-)
directions  by using density functional perturbation theory
(DFPT).

The second critical point $\epsilon_{c_{2}}$ is the yielding point
which is around $\epsilon=0.34$. The C-C distance corresponding to
$\epsilon_{c_{2}}$ is 2.02 \AA. Up to this point, the strain
energy always increases and the system preserves its
honeycomb-like structure. Upon the release of the tension, all the
deformation disappears and hence the system may return to its
original size at $\epsilon$=0. Furthermore, the value of
$\epsilon_{c_{2}}$ is found to depend on various defects and the
temperature of the system. For H-vacancy, we found that
$\epsilon_{c_{2}}$ is lowered to $\sim$0.21 as shown in
Fig.~\ref{fig:2} (c). As for C+H-vacancy, which corresponds to a
hole at one corner of hexagon, $\epsilon_{c_{2}}$ is further
lowered to 0.13 as shown in Fig.~\ref{fig:2} (d). We also examined
the effect of ambient temperature on the yielding strain.
Ab-initio molecular dynamic calculations (lasting 2 ps with time
steps of 2x10$^{-15}$ seconds) indicate that
$\epsilon_{c_{2}}$=0.34 corresponding to T=0 K is reduced to 0.20
at T=300 K and is further reduced to 0.18 at T=600 K. Apparently,
the yielding of perfect graphane under uniform strain at
$\epsilon_{c_{2}}$ can only occur for ideal conditions. For
$\epsilon_{c_{1}}<\epsilon<\epsilon_{c_{2}}$ the system is in a
meta-stable state. The long wave length perturbations, vacancy
defects, as well as high temperature effects lead
$\epsilon_{c_{2}}$ decrease to the strain values around
$\epsilon_{c_{1}}$. After the yielding point, where $\epsilon \geq
\epsilon_{c_{2}}$, the plastic range sets in with irreversible
deformations. This range, however, is beyond the scope of this
paper.

\begin{figure}
\includegraphics[width=8cm]{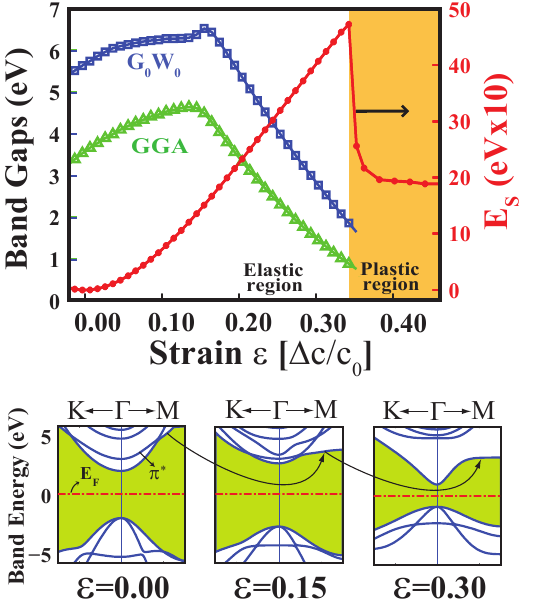}
\caption{(Color Online) The variation of energy band gaps with
(2D) uniform expansion. The band gaps obtained both from GGA
(green triangles) and G$_{o}$W$_{o}$ (blue squares) calculations
increase with increasing strain up to $\epsilon=0.15$, passes
through a maximum, then decrease until the yielding point. The
band gaps are given on the left and the strain energies are given
on the right. Three panels show how the bands at the edge of
conduction band are modified with strain.} \label{fig:3}
\end{figure}

We finally investigate the variation of the electronic properties
of graphane with  the uniform strain. The effect of strain on the
buckling is found to be minute. It decreases from 0.46 \AA~ to
0.43 \AA~ as $\epsilon$ increases from 0 to 0.30. Also, C-H bonds
are shortened only 1\% in this range of strain. The binding energy
of a single hydrogen in (10x10) supercell increases from 4.79 to
5.02 up to $\epsilon$=0.20. Normally, graphane is a semiconductor
with a wide direct band gap of 3.54 eV calculated by DFT-GGA, but
our calculations show that this gap can increase to 5.66 eV after
$G_{0}W_{0}$ corrections. On the other hand, recent $GW_{0}$ (5.97
eV)\cite{hasan} and $GW$ (5.4 eV)\cite{lebegue} corrections report
slightly different values depending on the method and parameters
used. More recently DFT-LDA calculations\cite{hakem} found the
band gap as 3.6 eV. Figure.~\ref{fig:3} shows the variation of GGA
and $G_{0}W_{0}$ band gap values with respect to the strain for
uniform expansion in the elastic region. While the lowest
conduction band is raised with strain in the first and second
panels; in the third panel, the second $\pi^*$ band is lowered
steadily and dips in the gap for $\epsilon > 0.15$. Dramatic
variation of the band gap with the strain suggests that graphane
can be used as a strain gauge at nanoscale.

In summary, we revealed the elastic constants of graphane
indicating that it has a quite high in-plane stiffness and very
low, perhaps the lowest Poisson's ratio among known monolayer
honeycomb structures. We showed that the band gap of graphane can
be modified significantly by applied strain in the elastic range.
It is suggested that elastic deformation can be used for further
functionalization of graphane and hence for monitoring its
chemical and electronic properties.

Part of the computations have been provided by UYBHM at Istanbul
Technical University through a Grant No. 2-024-2007.

\end{document}